\documentclass[prd,twocolumn,
showpacs,preprintnumbers,nofootinbib]{revtex4}
\usepackage[dvips]{graphicx}
\usepackage{enumerate}
\usepackage{amsmath,amssymb}
\usepackage{mathrsfs}
\usepackage[dvips]{graphicx}
\usepackage{bm}

\begin{document}
\preprint{CHIBA-EP-185/KEK Preprint 2010-22}

\title{Non-Abelian Dual Superconductor Picture for Quark Confinement 
}

\author{Kei-Ichi Kondo$^{1}$}
\email{kondok@faculty.chiba-u.jp}

\author{Akihiro Shibata$^{2}$}
\email{akihiro.shibata@kek.jp}

\author{Toru Shinohara$^{1}$}
\email{sinohara@graduate.chiba-u.jp}

\author{Seikou Kato$^{3}$}
\email{skato@fukui-nct.ac.jp}

\affiliation{$^1$Department of Physics,  
Graduate School of Science, 
Chiba University, Chiba 263-8522, Japan
\\
$^2$Computing Research Center,  High Energy Accelerator Research Organization (KEK),  Tsukuba  305-0801, Japan
and Graduate Univ. for Advanced Studies (Sokendai), Tsukuba  305-0801, Japan
\\
$^3$Fukui National College of Technology, Sabae 916-8507, Japan
}
\begin{abstract}

We give a theoretical framework for defining and extracting non-Abelian magnetic monopoles in a gauge-invariant way  in SU(N) Yang-Mills theory to study quark confinement. 
Then we give numerical evidences that the non-Abelian magnetic monopole defined in this way gives a dominant contribution to confinement of fundamental quarks in SU(3) Yang-Mills theory, which is in sharp contrast to the SU(2) case in which Abelian magnetic monopoles play the dominant role for quark confinement. 
\end{abstract}


\maketitle

\section{Introduction}
What is the mechanism for quark confinement?
The dual superconductor picture proposed long ago  \cite{dualsuper}  is believed to be a promising mechanics for quark confinement. 
For this mechanism to work, however, magnetic monopoles and their condensation are indispensable to cause the dual Meissner effect leading to the linear potential between quark and antiquark, namely,  
  area law of the Wilson loop average.

The Abelian projection method proposed by 't Hooft \cite{tHooft81,EI82} can be used to introduce such magnetic monopoles into the pure Yang-Mills theory even without matter fields, in sharp contrast to the 't Hooft--Polyakov magnetic monopole in the Georgi--Glashow gauge-Higgs model with adjoint matter fields. 
Indeed, numerical evidences supporting the dual superconductor picture resulting from such magnetic monopoles have been accumulated since 1990  in pure SU(2) Yang-Mills theory \cite{SY90,SNW94,AS99}.

However, {\it the Abelian projection method explicitly breaks both the local gauge symmetry and the global color symmetry} by partial gauge  fixing from an original non-Abelian gauge group $G=SU(N)$ to the maximal torus subgroup, $H=U(1)^{N-1}$. 
Moreover, the Abelian dominance \cite{SY90} and  magnetic monopole dominance \cite{SNW94} were observed only in a special class of gauges, e.g., the maximally Abelian (MA) gauge \cite{KLSW87} and Laplacian Abelian (LA) gauge, realizing the idea of  Abelian projection.

For $G=SU(2)$, we have already succeeded to settle the issue of  gauge (in)dependence by {\it introducing  a gauge-invariant magnetic monopole in a gauge independent way}, based on another method: a non-Abelian Stokes theorem for the Wilson loop operator \cite{DP89,Kondo98b} and a new reformulation of Yang-Mills theory rewritten in terms of new field variables \cite{KMS06,KMS05,Kondo06} and \cite{KKMSSI05,IKKMSS06,SKKMSI07}, elaborating the technique proposed by Cho \cite{Cho80} and Duan and Ge \cite{DG79} independently, and later readdressed by Faddeev and Niemi \cite{FN99} and Shabanov \cite{Shabanov99}.

For $G=SU(N)$, $N \ge 3$, there are no inevitable reasons why degrees of freedom associated with the maximal torus subgroup should be most dominant for quark confinement.
In this case, the problem is not settled yet.

In this paper, 
we give a theoretical framework for describing  {\it non-Abelian  dual superconductivity}, i.e., superconductivity caused by non-Abelian magnetic monopoles in $D$-dimensional $SU(N)$ Yang-Mills theory, which should be compared with the conventional Abelian  dual superconductivity, i.e., superconductivity caused by $U(1)^{N-1}$ magnetic monopoles in $SU(N)$ Yang-Mills theory, hypothesized by  Abelian projection. 
We demonstrate that {\it an effective low-energy description for quarks in the fundamental representation} {\it can be given by a set of non-Abelian restricted field variables} and that {\it non-Abelian $U(N-1)$ magnetic monopoles} in the sense of Goddard--Nuyts--Olive and Weinberg \cite{nAmm} {\it are the most dominant topological configurations for quark confinement} as conjectured in \cite{KT99,Kondo99Lattice99}.

This paper is organized as follows.

In section II, we rewrite the $SU(N)$ Wilson loop operator  in terms of a pair of gauge-invariant magnetic-monopole current  $k$ ($(D-3)$-form) and the associated geometric object $\Xi_{\Sigma}$ defined from the Wilson surface $\Sigma$ bounding the Wilson loop $C$, and another pair of an electric current $j$ (one-form  independently of $D$) and the associated topological object $N_{\Sigma}$, which follows from a non-Abelian Stokes theorem for the Wilson loop operator \cite{Kondo08}.   

In section III, we reformulate the $SU(N)$ Yang-Mills theory in terms of new field variables obtained by change of variables from the original Yang-Mills gauge field $\mathscr{A}_\mu^A(x)$ \cite{KSM08}, so that it gives an optimal description for the non-Abelian magnetic monopole defined from the $SU(N)$ Wilson loop operator in the fundamental representation of quarks.

In section IV, we construct a lattice version of the reformulated  Yang-Mills theory \cite{KSSMKI08,SKS09} to perform numerical simulations.  
The results of numerical simulations of the  lattice $SU(3)$ Yang-Mills theory give  numerical evidences that the restricted field variables become dominant in the infrared for correlation functions and the string tension ({\it infrared restricted non-Abelian dominance}) and that the $U(2)$ magnetic monopole gives a most dominant contribution to the string tension obtained from  $SU(3)$ Wilson loop average ({\it non-Abelian magnetic monopole dominance}). 
 
The final section is devoted to conclusion and discussion. 
We will mention a possible direction of future works needed to confirm the non-Abelian dual superconductivity.


\section{A non-Abelian Stokes theorem for the Wilson loop operator}

A version of a non-Abelian Stokes theorem (NAST) for the Wilson loop operator originally invented by Diakonov and Petrov \cite{DP89} for $G=SU(2)$ was proved to hold \cite{Kondo98b} and was extended to $G=SU(N)$ \cite{KT99,Kondo08} in a unified way \cite{Kondo08} as a path-integral representation by making use of a coherent state for the Lie group.  
For the Lie algebra $su(N)$-valued Yang-Mills field $\mathscr{A}_\mu(x)=\mathscr{A}_\mu^A(x) T_A$ with $su(N)$ generators $T_A$ ($A=1, \cdots, N^2-1$), 
the Wilson loop operator is defined by
\begin{align}
  W_C[\mathscr{A}]  
:=& {\rm tr} \left[ \mathscr{P} \exp \left\{ ig_{\rm YM}  \oint_{C} dx^\mu \mathscr{A}_\mu(x) \right\} \right]/{\rm tr}({\bf 1})  
 .
\end{align}
The NAST enables one to rewrite a non-Abelian Wilson loop operator $W_C[\mathscr{A}]$
 in terms of an Abelian-like potential $A_\mu$ as
\begin{equation}
 W_C[\mathscr{A}] = \int d\mu_{C}(g) \exp \left[  ig_{\rm YM} \oint_{C} A \right] ,
 \label{pre-NAST}
\end{equation}
where $g_{\rm YM}$ is the Yang-Mills coupling constant,  
  $d\mu_{C}(g):=\prod_{x \in C} d\mu(g_{x})$ with an invariant measure $d\mu$ on $G$  normalized as $\int d\mu(g_{x})=1$, $g_{x}$ is an element of a gauge group $G$ (more precisely, a representation $D_R(g_{x})$ of  $G$), and the one-form $A := A_\mu(x) dx^\mu$ is defined by
\begin{equation}
A_\mu(x) = {\rm tr}\{ \rho[ g_{x}^\dagger \mathscr{A}_\mu(x) g_{x} + ig_{\rm YM}^{-1} g_{x}^\dagger \partial_\mu g_{x} ] \} , \ g_{x} \in G .
\end{equation}
Here $\rho$ is defined as $\rho :=  | \Lambda \rangle \langle \Lambda |$ using a reference state (highest or lowest  weight state of the representation) $| \Lambda  \rangle$ by the use of a representation of the Wilson loop we consider. 
Note that ${\rm tr}(\rho) = \langle \Lambda | \Lambda \rangle = 1$ follows from the normalization of $| \Lambda \rangle$. 
Then it is rewritten into the surface-integral form using a usual Stokes theorem:
\begin{equation}
 W_C[\mathscr{A}] = \int d\mu_{\Sigma}(g) \exp \left[  ig_{\rm YM} \int_{\Sigma: \partial \Sigma=C} F \right] ,
\end{equation}
where $d\mu_{\Sigma}(g):=\prod_{x \in \Sigma} d\mu(g_{x})$, 
the two-form $F:=dA=\frac12 F_{\mu\nu}(x) dx^\mu \wedge dx^\nu$ is defined by
\begin{align}
  F_{\mu\nu}(x) &=  \sqrt{2(N-1)/N} [\mathscr{G}_{\mu\nu} (x)   
+ ig_{\rm YM}^{-1} {\rm tr} \{ \rho g_{x}^\dagger [\partial_\mu, \partial_\nu] g_{x} \} ],
\end{align}
with the field strength $\mathscr{G}_{\mu\nu}$ defined by
\begin{align}
 \mathscr{G}_{\mu\nu} (x) 
  &:=   \partial_\mu {\rm tr} \{ \bm{n}(x) \mathscr{A}_\nu(x) \} - \partial_\nu {\rm tr} \{ \bm{n}(x) \mathscr{A}_\mu(x) \} 
\nonumber\\& 
+ \frac{2(N-1)}{N} ig_{\rm YM}^{-1} {\rm tr} \{ \bm{n}(x) [\partial_\mu \bm{n}(x), \partial_\nu \bm{n}(x) ] \} 
 ,
\end{align}
and a normalized traceless field $\bm{n}(x)$ called the color field  
\begin{equation}
 \bm{n}(x) :=  \sqrt{N/[2(N-1)]} g_{x} \left[ \rho - \bm{1}/{\rm tr}(\bm{1}) \right] g_{x}^\dagger .
\end{equation}
Finally, the Wilson loop operator in the fundamental representation of $SU(N)$ is cast into the form \cite{Kondo08}:
\begin{align}
& W_C[\mathscr{A}] 
=  \int  d\mu_{\Sigma}(g)  \exp \left\{  ig_{\rm YM} (k, \Xi_{\Sigma}) + ig_{\rm YM} (j, N_{\Sigma}) \right\} ,
\label{NAST-SUN}
\nonumber\\
& k:=   \delta *f = *df, \quad j:=  \delta f , 
\quad
f:=  \sqrt{2(N-1)/N}  \mathscr{G}  ,
\nonumber\\
& \Xi_{\Sigma} :=  * d\Theta_{\Sigma} \Delta^{-1} = \delta *\Theta_{\Sigma} \Delta^{-1} , \
 N_{\Sigma} := \delta \Theta_{\Sigma} \Delta^{-1} ,
\end{align}
where two conserved currents,   ``magnetic-monopole current''   $k$ and  ``electric current'' $j$, are introduced, 
$\Delta:=d\delta+\delta d$ is the $D$-dimensional Laplacian in the $D$-dimensional Euclidean space, and $\Theta$ is an antisymmetric tensor of rank two  called the vorticity tensor:
$
 \Theta^{\mu\nu}_{\Sigma}(x) 
:=   \int_{\Sigma}  d^2S^{\mu\nu}(x(\sigma)) \delta^D(x-x(\sigma))  ,
$
which  has the support on the surface $\Sigma$ (with the surface element $dS^{\mu\nu}(x(\sigma))$) whose boundary is the loop $C$.
Incidentally, the last part $ig_{\rm YM}^{-1} {\rm tr} \{ \rho g_{x}^\dagger [\partial_\mu, \partial_\nu] g_{x} \}$ in $F$ corresponds to the Dirac string \cite{Kondo97,Kondo98a}, which is not gauge invariant and does not contribute to the Wilson loop in the end. 

For $SU(3)$ in the fundamental representation, the lowest-weight state $\langle \Lambda |=(0,0,1)$ leads to 
\begin{equation}
 \bm{n}(x) = g_{x} (\lambda_8/2) g_{x}^\dagger \in SU(3)/[SU(2) \times U(1)] \simeq CP^2 ,
\end{equation}
with the Gell-Mann matrix $\lambda_8:={\rm diag.}(1,1,-2)/\sqrt{3}$, 
while for $SU(2)$, $\langle \Lambda |=(0,1)$ yields
\begin{equation}
 \bm{n}(x) = g_{x} (\sigma_3/2) g_{x}^\dagger \in SU(2)/U(1) \simeq S^2 \simeq CP^1 ,
\end{equation}
with the Pauli matrix $\sigma_3:={\rm diag.}(1,-1)$.
The existence of magnetic monopole can be seen by a nontrivial Homotopy class of the map $\bm{n}$ from the sphere $S^2$ to the target space of the color field $\bm{n}$ \cite{KT99}:  
For $SU(3)$, 
\begin{align}
 & \pi_2(SU(3)/[SU(2) \times U(1)])=\pi_1(SU(2) \times U(1)) 
\nonumber\\&
=\pi_1(U(1))=\mathbb{Z} ,
\label{SU3-mm}
\end{align}
while for $SU(2)$
\begin{equation}
 \pi_2(SU(2)/U(1))=\pi_1(U(1))=\mathbb{Z} .
\end{equation}
For $SU(3)$, the magnetic charge of the non-Abelian magnetic monopole obeys the quantization condition \cite{Kondo08}:
\begin{equation}
 Q_m := \int d^3x k^0 = 2\pi \sqrt{3} g_{\rm YM}^{-1} n , \ n \in \mathbb{Z} .
\end{equation}
The NAST shows that {\it the $SU(3)$ Wilson loop operator in the fundamental representation detects the inherent $U(2)$ magnetic monopole which is  $SU(3)$ gauge invariant}, see (\ref{G-NF}). 
The representation can be classified by its {\it stability group} $\tilde H$ of $G$ \cite{KT99,Kondo08}.  
For the fundamental representation of $SU(3)$, the stability group  is $U(2)$.  
Therefore, the non-Abelian $U(2) \simeq SU(2)  \times U(1)$ magnetic monopole follows from the field in the representation with the stability group $\tilde H=SU(2)_{1,2,3} \times U(1)_{8}$, while the Abelian $U(1) \times U(1)$ magnetic monopole comes from that with $\tilde H=U(1)_{3} \times U(1)_{8}$. 
The adjoint representation belongs to the latter case. 
The former case occurs only when the weight vector of the representation is orthogonal to some of root vectors.  
The fundamental representation is indeed  this case. 
For $SU(2)$, such a difference does not exist and $U(1)$ magnetic monopoles appear irrespective of the representation, since $\tilde H$ is always $U(1)$ for any representation. 
For $SU(3)$, our result is different from Abelian projection in which two independent $U(1)$ magnetic monopoles appear for any representation, since 
\begin{align}
 & \pi_2(SU(3)/U(1) \times U(1))=\pi_1(U(1) \times U(1)) =\mathbb{Z}^2 .
\end{align}

\section{Reformulating the Yang-Mills theory using new variables}

Recently we have proposed a new reformulation \cite{KSM08} of the $SU(N)$  Yang-Mills (YM) theory based on new variables  by extending the Cho-Faddeev-Niemi (CFN) decomposition for $N \ge 3$ \cite{Cho80c,FN99a}. 
Our reformulation allows options discriminated by the stability group $\tilde{H}$ of the gauge group $G=SU(N)$. 
When $\tilde{H}$ agrees with the maximal torus group $H=U(1)^{N-1}$, it reproduces a manifestly gauge-independent reformulation of the  Abelian projection represented by the well-known maximal Abelian gauge.  This case is called the maximal option and agrees with the conventional CFNS decomposition for the $SU(N)$  Yang-Mills theory for $N \ge 3$ \cite{Cho80c,FN99a}. 
It was found \cite{KSM08} that there are the other options with the stability group $\tilde{H}$ other than the maximal torus group $H=U(1)^{N-1}$. 
Such possibilities are overlooked so far.
 Especially, the case of   $\tilde{H}=U(N-1)$ is called the minimal option. 
The minimal option gives the optimal description of quark in the fundamental representation combined with the non-Abelian Stokes theorem for the Wilson loop operator given above.

The reformulation enables one to understand quark confinement based on the dual superconductivity picture in a gauge independent way. 
This is because we can define gauge-invariant magnetic monopoles which are inherent in the Wilson loop operator.

In the realistic case of $SU(3)$, there are two options: the minimal option \cite{KSM08} with a single type of non-Abelian magnetic monopole characterized by the maximal stability subgroup $\tilde{H}=U(2)=SU(2)\times U(1)$, 
and the maximal one \cite{Cho80c,FN99a} with two types of Abelian magnetic monopoles characterized by the maximal torus subgroup $\tilde{H}=U(1)\times U(1)$.

We consider the decomposition of $\mathscr{A}_\mu(x)$ into two pieces $\mathscr{V}_\mu(x)$ and $\mathscr{X}_\mu(x)$:
\begin{equation}
 \mathscr{A}_\mu(x) = \mathscr{V}_\mu(x) + \mathscr{X}_\mu(x) ,
 \label{decomp}
\end{equation}
where the decomposed fields  $\mathscr V_\mu(x)$ and $\mathscr X_\mu(x)$ are  determined by solving the following defining equations (I) and (II), once a color field $\bm{n}(x)$ is given.
\\
\noindent
(I)  $\bm{n}(x)$ is a covariant constant in the background $\mathscr{V}_\mu(x)$:
\begin{align}
  0 = D_\mu[\mathscr{V}] \bm{n}(x) 
:=\partial_\mu \bm{n}(x) -  ig_{\rm YM} [\mathscr{V}_\mu(x), \bm{n}(x)]
 ,
\label{defVL2}
\end{align}
(II)  $\mathscr{X}^\mu(x)$  does not have the $\tilde{H}$-commutative part:
\begin{align}
  \mathscr{X}^\mu(x)_{\tilde{H}} := \left( {\bf 1} -   2\frac{N-1}{N}  [\bm{n} , [\bm{n} ,  \cdot ]]
\right) \mathscr{X}^\mu(x)   = 0  
\label{defXL2}
 . 
\end{align}

From the viewpoint of quark confinement, this decomposition has a remarkable property:  
It is shown \cite{KS} that (II) guarantees (a), while (I) guarantees (b). 
\\
(a)   $\mathscr{V}_\mu$ alone reproduces the Wilson loop operator: 
\begin{equation}
 W_C[\mathscr{A}] = W_C[\mathscr{V}] ,
 \label{W-dominant}
\end{equation}
where
\begin{align}
  W_C[\mathscr{V}]  
:=& {\rm tr} \left[ \mathscr{P} \exp \left\{ ig_{\rm YM}  \oint_{C} dx^\mu \mathscr{V}_\mu(x) \right\} \right]/{\rm tr}({\bf 1})  
 ,
\end{align}
(b) the field strength $\mathscr{F}_{\mu\nu}[\mathscr{V}] := \partial_\mu \mathscr{V}_\nu - \partial_\nu \mathscr{V}_\mu -ig_{\rm YM} [ \mathscr{V}_\mu,   \mathscr{V}_\nu ]$ in the color direction $\bm{n}$ agrees with $\mathscr{G}_{\mu\nu}$:
\begin{align}
 \mathscr{G}_{\mu\nu}(x) ={\rm tr} \{ \bm{n}(x) \mathscr{F}_{\mu\nu}[\mathscr{V}](x) \} .
 \label{G-NF}
\end{align}
This fact (a),(b) is also checked by using the explicit form of decomposed fields which are uniquely fixed by solving the defining equation: 
 
\begin{align}
 \mathscr{X}_\mu(x)  =& -ig_{\rm YM}^{-1}  \frac{2(N-1)}{N}  [\bm{n}(x) , \mathscr{D}_\mu[\mathscr{A}]\bm{n}(x)  ]
\in \mathcal{L}ie(G/\tilde{H}) ,
\nonumber\\
\mathscr V_\mu(x) 
 =& \mathscr C_\mu(x) 
  +\mathscr B_\mu(x) 
 \in  \mathcal{L}ie(\tilde{G}) 
 ,
\nonumber\\
  \mathscr{C}_\mu(x) 
:=& \mathscr{A}_\mu(x)  - \frac{2(N-1)}{N}   [\bm{n}(x) , [ \bm{n}(x) , \mathscr{A}_\mu(x) ] ]
\in  \mathcal{L}ie(\tilde{H}) 
,
\nonumber\\
 \mathscr{B}_\mu(x) 
:=& i g_{\rm YM}^{-1} \frac{2(N-1)}{N}[\bm{n}(x) , \partial_\mu  \bm{n}(x)  ] 
\in \mathcal{L}ie(G/\tilde{H}) 
 .
 \label{NLCV-minimal}
\end{align}
In what follows, $\mathcal{L}ie(G)$ denotes the Lie-algebra of the Lie group $G$.
\footnote{
If the color field $\bm{n}(x)$ is fixed to be a constant (this is considered to  be a gauge fixing), then we have $\mathscr{B}_\mu (x) \equiv 0$.  Then the decomposition 
$\mathscr A_\mu(x)=\mathscr V_\mu(x)+\mathscr X_\mu(x)$ 
is reduced to the orthogonal decomposition:  
$\mathscr{V}_\mu(x)=\mathscr{C}_\mu(x) = \mathscr{A}_\mu(x)  - \frac{2(N-1)}{N}   [\bm{n} , [ \bm{n} , \mathscr{A}_\mu(x) ] ] \in \mathcal{L}ie(\tilde{H})$ 
and 
$\mathscr{X}_\mu(x) =  \frac{2(N-1)}{N}   [\bm{n} , [ \bm{n} , \mathscr{A}_\mu(x) ] ]\in \mathcal{L}ie(G/\tilde{H})$.
In the $SU(2)$ case and the maximal option of the $SU(N)$ case, this procedure under the reduction condition (\ref{eq:diff-red}) is equivalent to taking the Maximal  Abelian gauge in the original Yang-Mills theory. But, the resulting theory in the minimal option of $SU(N)$ case has never been considered to the best of author's knowledge.  
This theory will be studied elsewhere. 
}

By combining (a) and (b) with the NAST given in the previous section, therefore, the Wilson loop operator can be rewritten   in terms of new variables:
\begin{align}
 W_C[\mathscr{A}] =& \int d\mu_{\Sigma}(g) 
 \exp \Big[  ig_{\rm YM} \sqrt{2(N-1)/N} 
\nonumber\\&
\quad\quad\quad\quad\quad\quad \times \int_{\Sigma: \partial \Sigma=C} {\rm tr} \{ \bm{n}  \mathscr{F}[\mathscr{V}]  \} \Big] .
\end{align}

In our reformulation, $\mathscr{V}_\mu(x)$ and $\mathscr{X}_\mu(x)$ must be  expressed in terms of $\mathscr{A}_\mu(x)$.
Therefore, we must give a procedure of determining $\bm{n}$ from  $\mathscr A_\mu$, thereby, all the new variables  $\mathscr{C}_\mu$, $\mathscr{X}_\mu$ and $\bm{n}$  are obtained 
from   $\mathscr{A}_\mu$ through the transformation law (\ref{NLCV-minimal}): 
\begin{equation}
 \mathscr{A}_\mu^A  \Longrightarrow (\bm{n}^\beta, \mathscr{C}_\nu^k,  \mathscr{X}_\nu^b)  
 .
\end{equation} 
To solve this issue, we begin with counting degrees of freedom:
$\mathscr A_\mu \in \mathcal{L}ie(G)=su(N)$ 
 means  
$\#[\mathscr A_\mu^A]=D  \cdot  {\rm dim}G=D(N^2-1)$, 
$\mathscr C_\mu \in \mathcal{L}ie(\tilde{H})=u(N-1)$
means 
$
\#[\mathscr C_\mu^k]=D  \cdot  {\rm dim}\tilde{H}=D(N-1)^2 
$ 
and 
$\mathscr X_\mu \in \mathcal{L}ie(G/\tilde{H})$ 
means
$
\#[\mathscr X_\mu^b]
=D  \cdot  {\rm dim}(G/\tilde{H})=2D(N-1)   
$ 
and
$\bm{n}  \in \mathcal{L}ie(G/\tilde{H})$ 
means
$\#[\bm{n}^\beta]= {\rm dim}(G/\tilde{H})=2(N-1)
$. 
Thus, the new variables $(\bm{n}^\beta, \mathscr{C}_\nu^k,  \mathscr{X}_\nu^b)$ have the $2(N-1)$ extra degrees of freedom, to be eliminated   to obtain the new theory equipollent to the original one. For this purpose, we impose $2(N-1)$ constraints $\bm\chi=0$, which we call the reduction condition.
For example, minimize the functional
\begin{align}
R[\mathscr{A}, \bm{n} ]
 :=   \int d^Dx \frac12 (D_\mu[\mathscr{A}]\bm{n})^2,
\end{align}
with respect to the enlarged gauge transformation: 
$
\delta\mathscr{A}_\mu=D_\mu[\mathscr{A}]\bm\omega,
$
and
$
\delta \bm{n}
 =gi[\bm\theta , \bm{n} ]
 =gi[ \bm\theta_\perp , \bm{n} ]
$
where 
$\bm\omega \in \mathcal{L}ie(G)$ and $\bm\theta_\perp \in \mathcal{L}ie(G/\tilde{H})$.
Then, we find 
$
\delta R[\mathscr{A}, \bm{n}]
=g \int d^Dx
 (\bm\theta_\perp-\bm\omega_\perp)
 \cdot i[ \bm{n} , D^\mu[\mathscr{A}]D_\mu[\mathscr{A}]\bm{n} ]
 ,
$
where $\bm\omega_\perp$ denotes the component of $\bm\omega$ in the direction $\mathcal{L}ie(G/\tilde{H})$. 
The minimization  $\delta R[\mathscr{A}, \bm{n}]=0$ imposes no condition for $\bm\omega_\perp = \bm\theta_\perp$ (diagonal part of $G \times G/\tilde{H}$), while 
\begin{equation}
\bm\chi[\mathscr{A},\bm{n}]
 :=[ \bm{n} ,  D^\mu[\mathscr{A}]D_\mu[\mathscr{A}]\bm{n} ]
 = 0 
  ,
\label{eq:diff-red}
\end{equation}
is imposed for $\bm\omega_\perp   \not=  \bm\theta_\perp$ (off-diagonal part of $G \times G/\tilde{H}$).
The number of constraint is 
$
\#[\bm\chi]= {\rm dim}(G \times G/\tilde{H})- {\rm dim}(G)= {\rm dim}(G/\tilde{H})=2(N-1)=\#[\bm{h}^\beta]
$ as desired. 
As a bonus, the color field $\bm{n}(x)$ is determined by solving (\ref{eq:diff-red}) for  given   $\mathscr{A}_\mu(x)$.
This completes the procedure.

The Wilson loop average $W_{C}$ is defined by 
\begin{align}
 W_{C} 
= \langle W_C[\mathscr{A}] \rangle_{\rm YM}
:=  Z_{{\rm YM}}^{-1} \int \mathcal{D}\mathscr{A}_\mu^A e^{-S_{{\rm YM}}[\mathscr{A}]} W_C[\mathscr{A}] ,
\end{align}
with  the partition function 
$
Z_{{\rm YM}} =  \int \mathcal{D}\mathscr{A}_\mu^A e^{-S_{{\rm YM}}[\mathscr{A}]} 
$
by omitting the gauge fixing to simplify the expression. 
The pre-NAST (\ref{pre-NAST}) tells us that
\begin{equation}
 W_{C} =  Z_{{\rm YM}}^{-1} \int  d\mu_{C}(g)   \mathcal{D}\mathscr{A}_\mu^A e^{-S_{{\rm YM}}[\mathscr{A}]} 
   e^{  ig_{\rm YM} \oint_{C} A } .
\end{equation}
Inserting  
$
1 = \int \mathcal{D}n^\alpha \prod_{x}  \delta(\bm{n}(x)-g_{x}(\lambda_8/2)g^\dagger_{x})
$ 
yields 
\begin{align}
 W_{C} =&  Z_{{\rm YM}}^{-1} \int  d\mu_{C}(g) \int \mathcal{D}\mathscr{A}_\mu^A  \mathcal{D}n^\alpha \delta(\bm{n}(x)-g_{x}(\lambda_8/2)g^\dagger_{x})
 \nonumber\\&
 \times   e^{-S_{{\rm YM}}[\mathscr{A}]}  e^{  ig_{\rm YM} \oint_{C} A  } 
.
\end{align}
Thus, in the reformulated theory in which $n^\beta(x)$, $\mathscr{C}_\nu^k(x)$, $\mathscr{X}_\nu^b(x)$ are {\it independent} field variables,  $W_{C}$ is written
\begin{align}
 W_{C} =&  \tilde{Z}_{{\rm YM}}^{-1} \int  d\mu_{\Sigma}(g) \int  \mathcal{D}\mathscr{C}_\nu^k \mathcal{D}\mathscr{X}_\nu^b  \mathcal{D}n^\beta
\delta(\tilde{\bm\chi}) \Delta_{\rm FP}^{\rm red}
\tilde{J}
 \nonumber\\&
 \times   e^{-\tilde S_{\rm YM}[\bm n, \mathscr{C},\mathscr{X}]}  e^{  ig_{\rm YM} (k, \Xi_{\Sigma}) + ig_{\rm YM} (j, N_{\Sigma}) }  
 \nonumber\\ 
=& \langle e^{  ig_{\rm YM} (k, \Xi_{\Sigma}) + ig_{\rm YM} (j, N_{\Sigma}) }    \rangle_{\rm YM}
 ,
 \label{Wkj}
\end{align}
where 
the Yang-Mills action is rewritten in terms of new variables using (\ref{decomp}) and (\ref{NLCV-minimal}), 
$
\tilde S_{\rm YM}[\bm n, \mathscr{C},\mathscr{X}]=S_{\rm YM}[\mathscr A]
$
and the new partition function is introduced:
$
 \tilde{Z}_{{\rm YM}}  
= \int \mathcal{D}\mathscr{C}_\nu^k \mathcal{D}\mathscr{X}_\nu^b  \mathcal{D}n^\beta
\delta(\tilde{\bm\chi}) \Delta_{\rm FP}^{\rm red} \tilde{J}
 e^{-\tilde S_{\rm YM}[\bm n, \mathscr{C},\mathscr{X}]} 
$.
It is shown \cite{KSM08} that the integration measure $\mathcal{D}\mathscr{A}_\mu^A$  is finally transformed to 
$
 \mathcal{D}\mathscr{C}_\nu^k \mathcal{D}\mathscr{X}_\nu^b \mathcal{D}n^\beta 
\delta(\tilde{\bm\chi}) \Delta_{\rm FP}^{\rm red} \tilde{J}
$, 
where 
(i) the Jacobian  $\tilde{J}$ 
is very simple, 
$
  \tilde{J} = 1 ,
$
\cite{KSM08} 
irrespective of the choice of reduction condition,  
(ii)  $\bm{\chi}[ \mathscr{A},\bm{n}]=0$ is rewritten in terms of new  variables:
$
\tilde{\bm\chi} 
 :=\tilde{\bm\chi} [\bm n, \mathscr{C},\mathscr{X}]
 :=D^\mu[\mathscr{V}]\mathscr{X}_\mu 
 , 
$
and 
(iii) the associated Faddeev-Popov determinant  $\Delta_{\rm FP}^{\rm red}$   
is calculable   using the BRST method, e.g.\cite{KMS05}.

In the previous section, the Wilson loop operator has been exactly rewritten in terms of the gauge-invariant magnetic current $k$ (and the electric current $j$).  This shows that the Wilson loop operator can be regarded as a probe of magnetic monopoles.
In this section, moreover, we  have succeeded to connect the Wilson loop  average  with magnetic monopoles which is supposed to be a basic ingredient to cause dual superconductivity as a promising mechanism of quark confinement. 
In fact, Eq.(\ref{Wkj}) tells us what quantity we should examine to see the magnetic monopole contribution to the Wilson loop average. This equation is important to give a connection between our formulation and magnetic monopole inherent in the Wilson loop to see quark confinement. 
In fact, we give numerical calculations of the potential $V_m(R)$ in the next section based on the lattice version of  Eq.(\ref{Wkj}).

\section{Lattice SU(N) Yang-Mills theory} 

\subsection{Reformulating the lattice $SU(N)$ Yang-Mills theory}

We have reformulated the Yang-Mills theory with a gauge group $G=SU(N)$ also on a lattice using new variables in the same spirit as in the continuum, see \cite{KSSMKI08,SKS09} for the details. 
But we just summarize it below. 

First, gauge field configurations  $\{ U_{x,\mu} \}$ on a four-dimensional Euclidean lattice are generated by using the standard method: the Wilson action and pseudo heat-bath algorithm.  
The gauge variable $U_{x,\mu}$ on the link has the gauge transformation: 
\begin{equation}
  U_{x,\mu}  \rightarrow \Omega_{x} U_{x,\mu} \Omega_{x+\mu}^{-1} = U_{x,\mu}^\prime 
  , \quad \Omega_{x} \in G
   .
   \label{U-transf}
\end{equation}

Second, according to the  continuum formulation \cite{KSM08}, we introduce just a single color field $\bm{n}_{x}$ even for  $G=SU(N)$ ($N \ge 2$)  in the minimal option.   
The color field $\bm{n}_{x}$ on a lattice is regarded as a site variable defined on a site $x$ and takes the value as
\begin{equation}
 \bm{n}_{x} = \bm{n}_{x}^A T_A  \in G/\tilde{H} =SU(N)/U(N-1) \simeq CP^{N-1} ,
\end{equation}
with a unit length
\begin{equation}
  \bm{n}_{x}^A  \bm{n}_{x}^A  = 1 .
\end{equation}
For a given set of gauge field configurations $\{ U_{x,\mu} \}$,  a set of color fields $\{ \bm{n}_{x} \}$ is determined by imposing a lattice version of the reduction condition.  
A reduction condition in the minimal option on a lattice is given  by minimizing the reduction functional $F_{\rm red}$ for  a given set of gauge field configurations  $\{ U_{x,\mu} \}$ with respect to the color field $\{ \bm{n}_{x} \}$:
\begin{align}
 F_{\rm red}[\bm{n},U] 
:=& \epsilon^{D} \sum_{x,\mu} {\rm tr}\{ (D_\mu^{\epsilon}[U]\bm{n}_{x}) (D_\mu^{\epsilon}[U]\bm{n}_{x})^\dagger \}/{\rm tr}(\bf{1}) 
\nonumber\\
=& \epsilon^{D-2} \sum_{x,\mu}[1- 2{\rm tr}( \bm{n}_{x} U_{x,\mu} \bm{n}_{x+\mu} U_{x,\mu}^\dagger ) ]
 ,   
\end{align}
where $D_\mu^{\epsilon}[U]$ is the lattice covariant derivative in the adjoint representation defined by 
$
 D_\mu^{\epsilon}[U] \bm{n}_{x} := \epsilon^{-1}( U_{x,\mu} \bm{n}_{x+\mu} - \bm{n}_{x} U_{x,\mu}) 
$
with a lattice spacing $\epsilon$.
Thus, a set of color fields $\bm{n}(x)$  we need is obtained as a set of unit vector fields $\tilde{\bm{n}}(x)$ which realizes the minimum of the reduction functional:
\begin{align}
 F_{\rm red}[\bm{n},U] 
 =&   \min_{\tilde{\bm{n}}} F_{\rm red}[\tilde{\bm{n}},U]
 ,   
\end{align}
It is observed that solving the reduction problem is equivalent to finding the ground state of the spin-glass model, since the reduction functional $F_{\rm red}$ is minimized with respect to the color field $\{ \bm{n}_{x} \}$ under the random link interaction $J^{AB}_{x,\mu}[U]$ for given gauge field configurations  $\{ U_{x,\mu} \}$:
\begin{align}
 F_{\rm red}[\bm{n},U] 
 =&  \epsilon^{D-2} \sum_{x,\mu}(1-  J^{AB}_{x,\mu}[U]  \bm{n}_{x}^A  \bm{n}_{x+\mu}^B ) ,
\nonumber\\
 J^{AB}_{x,\mu}[U] :=& 2{\rm tr}( T_A U_{x,\mu} T_B U_{x,\mu}^\dagger ) 
 .   
\end{align}
This observation has been actually used to find the minimum in SU(2) case \cite{Kato-lattice2009}.
After applying the reduction condition  \cite{KSSMKI08},   the color field  transforms under the gauge transformation in the adjoint way: 
\begin{equation}
  \bm{n}_{x}  \rightarrow \Omega_{x} \bm{n}_{x} \Omega_{x}^{-1} = \bm{n}_{x}^\prime
  , \quad \Omega_{x} \in G
  .
  \label{n-transf}
\end{equation}
The reduction functional $F_{\rm red}$ is invariant under the gauge transformation. Therefore, imposing the reduction condition does not break the original gauge invariance. 
We can impose any gauge fixing afterwards, if necessary. 
The details for the algorithm of the reduction procedure on a lattice in the SU(3) case will be given in \cite{SKKS11}.

Third, new variables on a lattice are introduced by using the lattice version of  change of variables \cite{KSSMKI08,SKS09}: Once a set of color fields $\bm{n}_{x}$ is given,  the $G=SU(N)$-valued gauge variable $U_{x,\mu} \in G$ is decomposed into the product of two $G$-valued variables $X_{x,\mu}$ and $V_{x,\mu}$ defined on the same lattice:
\begin{equation}
 U_{x,\mu} = X_{x,\mu} V_{x,\mu} \in G ,
 \quad X_{x,\mu}, V_{x,\mu} \in G
  ,
  \label{decomp-1}
\end{equation}
where  the lattice variables $V_{x,\mu}$ and $X_{x,\mu}$ are supposed to be related to the Lie-algebra $\mathscr{V}_\mu(x)$ and  $\mathscr{X}_\mu(x)$ as  
\begin{equation}
  V_{x,\mu} = \exp \{-i\epsilon g_{\rm YM} \mathscr{V}_\mu(x) \}  , \ 
  X_{x,\mu} = \exp \{-i\epsilon g_{\rm YM} \mathscr{X}_\mu(x) \}   
  ,
\end{equation}
just as
\begin{equation}
  U_{x,\mu} = \exp \{ -i\epsilon g_{\rm YM} \mathscr{A}_\mu(x) \} 
   .
\end{equation}

We require that $V_{x,\mu}$ is a new link variable which transforms like a usual gauge variable $U_{x,\mu}$ on the same  link: 
\begin{equation}
  V_{x,\mu}  \rightarrow \Omega_{x} V_{x,\mu} \Omega_{x+\mu}^{-1} = V_{x,\mu}^\prime 
  , \quad \Omega_{x} \in G
   .
   \label{V-transf}
\end{equation}
For this gauge transformation to be consistent with the decomposition (\ref{decomp-1}), consequently, $X_{x,\mu}$ must behave like an adjoint matter field defined at the site $x$ under the gauge transformation:
\begin{equation}
  X_{x,\mu}
 \rightarrow \Omega_{x} X_{x,\mu} \Omega_{x}^{-1} = X_{x,\mu}^\prime
  , \quad \Omega_{x} \in G
  .
  \label{X-transf1}
\end{equation}
These properties of the decomposed variables under the gauge transformation are expected from the continuum version. 
The decomposed variables $X_{x,\mu}$ and $V_{x,\mu}$ are determined by solving defining equations. 
A lattice version of the first defining equation proposed in \cite{KSSMKI08,SKS09} is:
(I)~The color field $\bm{n}_{x}$ is covariantly constant in the (matrix) background $V_{x,\mu}$:
\begin{equation}
  0 = \epsilon D_\mu^{(\epsilon)}[V]\bm{n}_{x} := V_{x,\mu} \bm{n}_{x+\mu} - \bm{n}_{x} V_{x,\mu}   ,
  \label{lat-defeq-min1}
\end{equation}
where $D_\mu^{(\epsilon)}[V]$ is the lattice covariant derivative in the adjoint representation. 
The solution of this defining equation can be obtained exactly for any $N$ \cite{SKS09} (without using the ansatz employed in \cite{KSSMKI08} to find the solution for $N=2,3$) to give $X_{x,\mu}$ and $V_{x,\mu}=X_{x,\mu}^\dagger U_{x,\mu}$:
\begin{subequations}
\begin{align}
   X_{x,\mu}
=& \hat{L}_{x,\mu}^\dagger (\det (\hat{L}_{x,\mu}))^{1/N} g_{x}^{-1} 
 ,
\label{eq:Xmin}
\\
V_{x,\mu} 
=&  g_{x} \hat{L}_{x,\mu} U_{x,\mu} (\det (\hat{L}_{x,\mu}))^{-1/N}   
   ,
\label{eq:KXlc2min}
\end{align}
where
\begin{align}
 \hat{L}_{x,\mu}
=& (\sqrt{L_{x,\mu}L_{x,\mu} ^\dagger})^{-1} L_{x,\mu} ,
\nonumber\\ 
\hat{L}_{x,\mu}^\dagger
=& L_{x,\mu}^\dagger \left( \sqrt{L_{x,\mu} L_{x,\mu}^\dagger}\right)^{-1} 
 ,
\end{align} 
with
\begin{align}
L_{x,\mu}  
  =& \frac{N^{2}-2N+2}{N}\mathbf{1}
\nonumber\\&
+\left(  N-2\right)  \sqrt{\frac{2(N-1)}{N} 
}\left(  \bm{n}_{x}+U_{x,\mu}\bm{n}_{x+\mu}U_{x,\mu}^{-1}\right)
\nonumber\\&
 +4\left(  N-1\right)  \bm{n}_{x}U_{x,\mu}\bm{n}_{x+\mu}U_{x,\mu}%
^{-1} .
\end{align}
\end{subequations}
Here  a common factor $g_{x}$  in the above expressions for $X_{x,\mu}$ and $V_{x,\mu}$ is the part undetermined from the first defining equation alone. 
In fact, $g_{x}$ is an element of the extra  symmetry associated with the decomposition \cite{SKS09}: $Z(N) \times \tilde{H}$,  $\tilde{H} = U(N-1) \subset SU(N)$:
\begin{align}
   g_{x}
=& e^{ -2\pi i q_{x}/N }  \exp \left\{  - ia_{x} \bm{n}_{x} - i \sum_{\ell=1}^{(N-1)^2-1} a_{x}^{(\ell)} \mathbf{u}_{x}^{(\ell)} \right\} 
\nonumber\\&
 (q_{x}=0, \cdots, N-1) ,
\end{align}
where $a_{x},  a_{x}^{(\ell)} \in \mathbb{R}$ and $\{ \bm{u}_{x}^{(\ell)} \}$ is a set of  Hermitian traceless generators of $SU(N-1)$ commutable with $\bm{n}_{x}$.

In order to fix it, we must impose further conditions.  Hence we impose  the second defining equation, e.g., 
(II)~$g_{x}$ is equated with an element $g_{x}^{0}$:
\begin{equation}
  g_{x}=g_{x}^{0} .
\end{equation}
The simplest one is to take $g_{x}^{0}=\mathbf{1}$. 
Thus the decomposed variables $X_{x,\mu}$ and $V_{x,\mu}$ are completely determined. 
It can be checked that the lattice formulation given in this section reproduces the continuum formulation given in the previous section, in the naive continuum limit $\epsilon \rightarrow 0$.

\subsection{Numerical simulations for $SU(3)$ Yang-Mills theory} 

In the new formulation, we can define another non-Abelian Wilson loop operator $W_C[\mathscr{V}]$ by replacing $\mathscr{A}$ by $\mathscr{V}$ in the original definition of the Wilson loop operator $W_C[\mathscr{A}]$. 
For the lattice version of the Wilson loop operator $W_C[\mathscr{A}]$,
\begin{align}
  W_C[U]  
:=  {\rm tr} \left[ \prod_{<x,x+\mu> \in C}  U_{x,x+\mu}  \right]/{\rm tr}({\bf 1})  
 ,
\end{align}
the lattice version of $W_C[\mathscr{V}]$ is easily constructed:
\begin{align}
  W_C[V]  
:=  {\rm tr} \left[ \prod_{<x,x+\mu> \in C}  V_{x,x+\mu}  \right]/{\rm tr}({\bf 1})  
 .
\end{align}
This is invariant under the gauge transformation (\ref{V-transf}). 
Moreover, we define  the lattice version $K$ of the magnetic monopole current $k$ defined in (\ref{NAST-SUN}):
\begin{align}
   K_{x,\mu} 
:=&   \partial_\nu {}^* \Theta_{x,\mu\nu}
= \frac12 \epsilon_{\mu\nu\alpha\beta} \partial_\nu \Theta_{x,\alpha\beta}
, 
  \nonumber\\
 \epsilon^2 \Theta_{x,\alpha\beta} 
:=&   {\rm arg} \Big[ {\rm tr} \Big\{ \left( \frac13 \bm{1} - \frac{2}{\sqrt{3}} \bm{n}_{x} \right) 
\nonumber\\&
\times V_{x,\alpha}V_{x+\alpha,\beta}V_{x+\beta,\alpha}^\dagger V_{x,\beta}^\dagger \Big\} \Big] 
  .
\end{align}
It is easy to observe that $\Theta_{x,\mu\nu}$ is invariant under the gauge transformation (\ref{n-transf}) and (\ref{V-transf}),  and hence $K_{x,\mu}$ is also gauge-invariant. 
Then we can define the magnetic-monopole part of the Wilson loop operator by
\begin{align}
W_C[K] :=&   \exp \left( i  \sum_{x,\mu} K_{x,\mu} \Xi^{\Sigma}_{x,\mu} \right)
    ,
\nonumber\\
 \Xi^{\Sigma}_{x,\mu} :=& {\displaystyle\sum_{s^{\prime}}}
 \Delta_L^{-1}(s-s^{\prime})\frac{1}{2}\epsilon_{\mu\alpha\beta\gamma} 
\partial_{\alpha}S_{\beta\gamma}^{J}(s^{\prime}+\mu) 
 , 
 \label{WK}
\end{align}
where $S_{\beta\gamma}^{J}(s^{\prime}+\mu)$ is a plaquette variable satisfying $\partial^{'}_{\alpha} S_{\alpha\beta}^{J}(x)=J_{\beta}(x)$ with the external source $J_{x,\mu}$ introduced to calculate the static potential, 
$\partial'$ denotes the backward lattice derivative
$\partial_{\mu}^{'}f_x=f_x-f_{x-\mu}$,  
$S^J_{x,\beta\gamma}$ denotes a surface bounded by the closed loop $C$ on which the electric source $J_{x,\mu}$ has its support, 
and $\Delta_L^{-1}(x-x')$ is the inverse Lattice Laplacian, see e.g., \cite{CFKMPS00}.

Numerical simulations are performed for SU(3) Yang-Mills theory on the $24^4$ lattice according to the lattice reformulation explained above for $N=3$. 
More details of numerical simulations will be given in a subsequent paper \cite{SKKS11}. 

The static quark-antiquark potential $V_f(R)$ is defined by taking the limit $T \rightarrow \infty$ from the Wilson loop average $\langle W_C[U]  \rangle$ for a rectangular loop $C=R \times T$:
\begin{align}
  V_f(R) 
=& - \lim_{T \rightarrow \infty} \frac{1}{T} \ln 
\langle W_C[U] \rangle
 . 
 \label{VfR}
\end{align}
In practice  \cite{Shibata-lattice2010},  we fit  numerical data of $\langle W_C[U] \rangle$ by the two-variable function $W(R,T)$ according to 
\begin{subequations}
\begin{align}
  \langle W_C[U] \rangle =& \exp (-W(R,T) ),
 \label{WVa}
\\
 W(R,T) :=& T V(R) + (a_1 R+b_1+c_1/R) 
\nonumber\\&
+ (a_2 R+b_2+c_2/R) T^{-1}   ,
 \label{WVb}
\\
 V(R) :=& \sigma R + b + c/R 
 , 
 \label{WVc}
\end{align}
\label{WV}
\end{subequations}
and determine all coefficients in $W(R,T)$. 
Then we identify  $V_f(R)$ with $V(R)$ to be obtained by extrapolating $W(R,T)/T$ to $T \rightarrow \infty$:
\begin{align}
  V(R) 
= \lim_{T \rightarrow \infty}  \frac{W(R,T)}{T}   
= V_f(R)
 . 
 \label{Vf}
\end{align}
Here the coefficient $\sigma$ of the linear part of the potential (\ref{WVc}) is the string tension which equals to  the slope of the curve for large $R$.

\begin{figure}[t]
\begin{center}
\includegraphics[height=5.0cm,width=8.0cm]{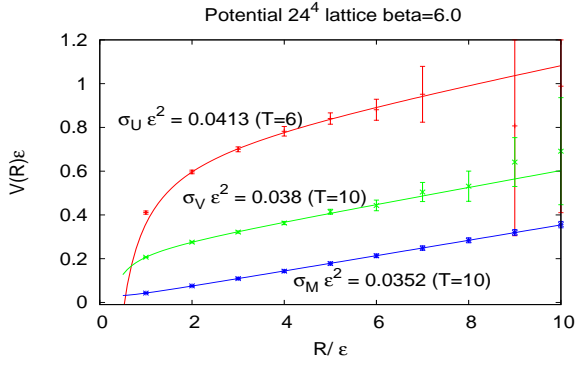}
\end{center}
\caption{$SU(3)$ quark-antiquark potentials as functions of the quark-antiquark distance $R$: (from above to below) 
 (i) full potential $V_f(R)$ (red curve),  (ii) restricted part $V_r(R)$ (green curve) and (iii) magnetic--monopole part  $V_m(R)$ (blue curve), measured at $\beta=6.0$ on $24^4$ using 500 configurations where $\epsilon$ is the lattice spacing.}
\label{fig:quark-potential}
\end{figure}

In Fig. \ref{fig:quark-potential}, we compare the three quark-antiquark potentials (i), (ii) and (iii).  For each potential, we plot a set of point data for a specified value of $T$ (e.g., $T=6, 10$):
\begin{align}
    - \frac{1}{T} \ln \langle W_{C}[\cdot] \rangle \quad\text{versus} \quad R 
 , 
 \label{VT}
\end{align}
and the curve 
\begin{align}
    V(R) =& \sigma R + b + c/R ,
\end{align}
extrapolated to $T \rightarrow \infty$ according to (\ref{WV}) and (\ref{Vf}): 
\begin{enumerate}
\item
[(i)] the full potential $V_f(R)$  calculated from the standard  $SU(3)$ Wilson loop average 
$\langle W_C[U] \rangle$,

\item
[(ii)] the restricted potential $V_r(R)$ calculated from the decomposed variable $\mathscr{V}$ through the restricted Wilson loop average $\langle W_C[V] \rangle$
\begin{align}
  V_r(R) 
=& - \lim_{T \rightarrow \infty} \frac{1}{T} \ln 
\langle W_C[V] \rangle
 , 
 \label{Vr}
\end{align}

\item
[(iii)] magnetic--monopole contribution  $V_m(R)$ calculated from the lattice counterpart (\ref{WK}) of the continuum quantity   $\langle e^{  i (k, \Xi_{\Sigma})  }  \rangle$ according to (\ref{Wkj}):
\begin{align}
V_m(R) 
=& - \lim_{T \rightarrow \infty} \frac{1}{T} \ln  \langle W_C[K] \rangle 
    ,
\end{align}

\end{enumerate}
Three potentials are gauge invariant quantities by construction.

The results of our numerical simulations exhibit infrared restricted variable $\mathscr{V}$ dominance in the string tension, e.g.,
\begin{equation}
\frac{\sigma_V}{\sigma_U} = \frac{0.0380}{0.0413} \simeq 0.92,
\end{equation}
and  non-Abelian  magnetic monopole dominance in the string tension, e.g.,
\begin{equation}
\frac{\sigma_M}{\sigma_U} = \frac{0.0352}{0.0413} \simeq 0.85 .
\end{equation}
However, we know that $\sigma_U$ has the largest errors among three string tensions.
Incidentally, if we use the other data for $\epsilon \sqrt{\sigma_U*}$  at $\beta=6.0$ given in Table 4 of \cite{EHK98} where 
$\epsilon^2 \sigma_U* = (\epsilon \sqrt{\sigma_U*})^2 = 0.2154^2 \sim 0.2209^2=0.0464 \sim 0.0488$, 
the ratios of two string tensions $\sigma_V, \sigma_M$ to the total string tension $\sigma_U$ are modified  
\begin{align}
\frac{\sigma_V}{\sigma_U*} & \cong 0.78  \sim  0.82,
\\
\frac{\sigma_M}{\sigma_U*} & \cong 0.72  \sim  0.76 ,
\end{align}
Anyway, we have obtained the infrared restricted variable $\mathscr{V}$ dominance in the string tension (78--82\%) and the non-Abelian  magnetic monopole dominance in the string tension  (72--76\%).
Both dominance are obtained in the gauge independent way. 
\footnote{The method of fitting the data given in this paper is the same as that in \cite{Shibata-lattice2010}, but is different from that used in \cite{Shibata-lattice2008}. 
}

To obtain correlation functions of field variables, we need to fix the gauge and we have adopted the Landau gauge for the original Yang-Mills field $\mathscr{A}$ so that the global color symmetry is not broken.  This property is desirable to study color confinement, but it is lost in the MA gauge.

\begin{figure}[t]
\begin{center}
\includegraphics[height=3.0cm,width=4.0cm]{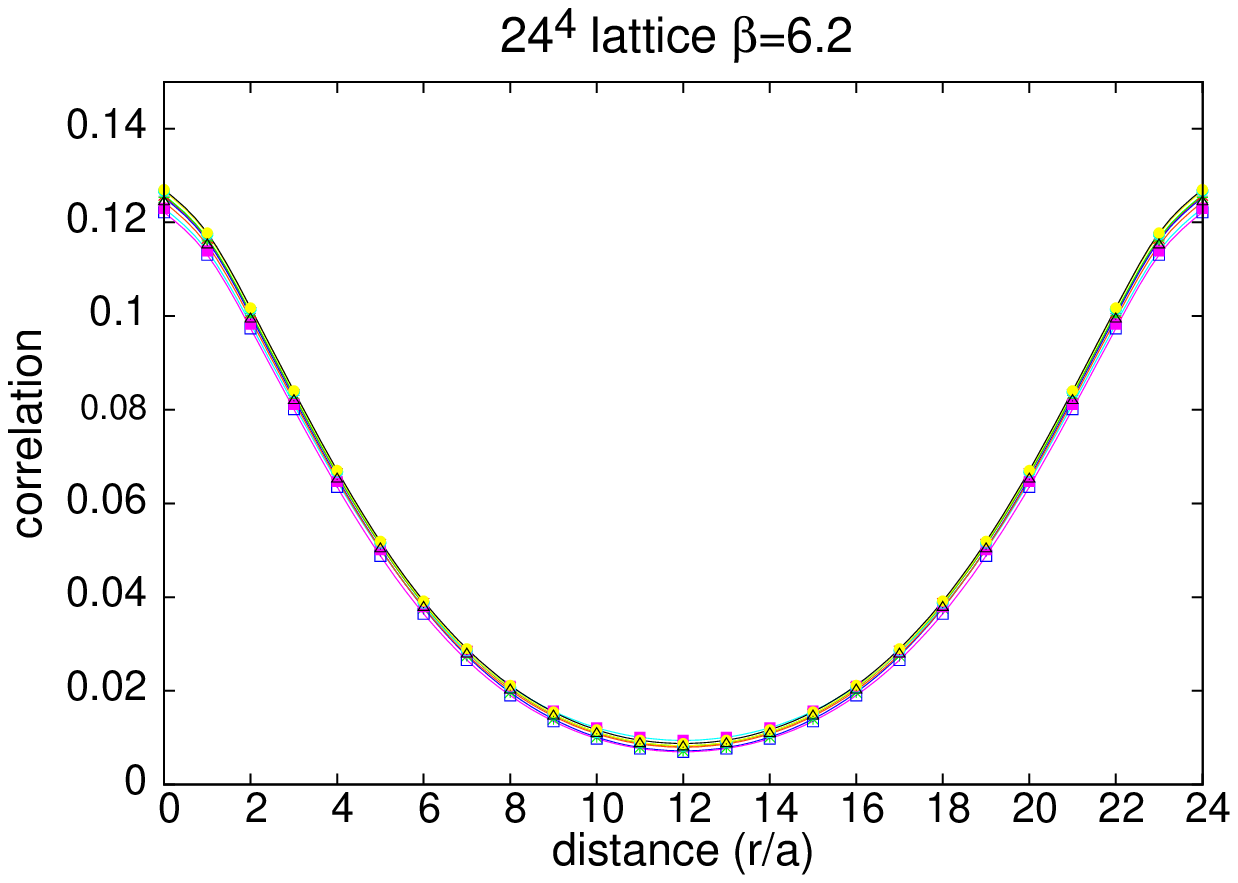}
\includegraphics[height=3.0cm,width=4.0cm]{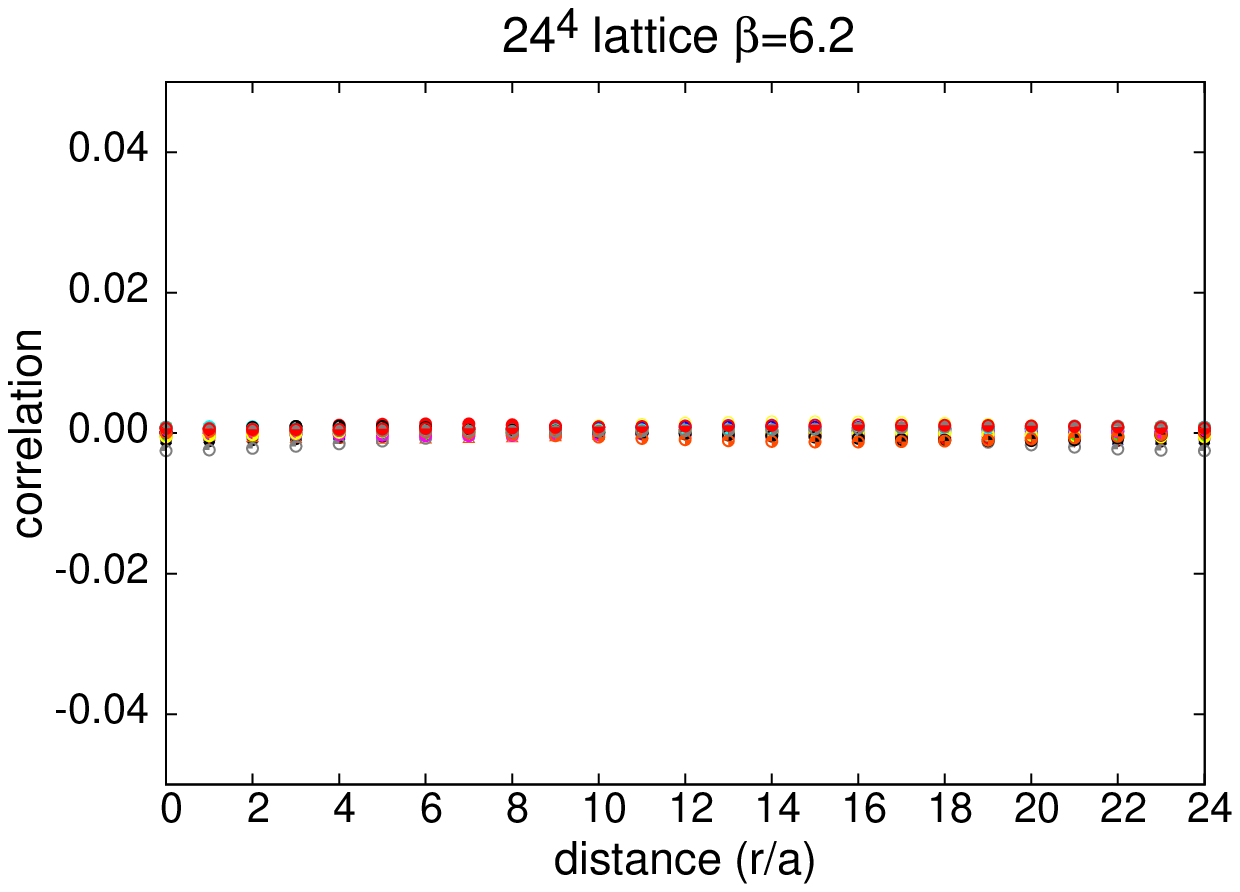}
\end{center}
\caption{Color field correlators $\langle  n^A(x) n^B(0) \rangle$ ($A,B=1, \cdots, 8$) as functions of the distance $r:=|x|$ measured at  $\beta=6.2$ on   24$^4$ lattice,  using 500 configurations under the Landau gauge. 
(Left panel) $A=B$, 
(Right panel) $A \not= B$.
}
\label{fig:color-field-corr}
\end{figure}

Fig.\ref{fig:color-field-corr} shows  two-point correlation functions of color field $\langle  n^A(x) n^B(0) \rangle$ versus the distance $r:=|x|$. 
All plots of correlators for $A=B=1,2, \cdots, 8$ overlap on top of each other, and hence they can be fitted by a common non-vanishing  function $D(r)$ (left panel), while all correlators for $A \not= B$ are nearly equal to zero (right panel). 
Therefore,  the correlators $\langle  n^A(x) n^B(0) \rangle$ are of the form:
\begin{equation}
\langle  n^A(x) n^B(0) \rangle = \delta^{AB} D(r) \quad (A,B=1,2, \cdots, 8)
 .
\end{equation}
We have also checked that  one-point functions vanish, 
\begin{equation}
\langle  n^A(x)  \rangle = \pm 0.002 \simeq 0 \quad (A =1,2, \cdots, 8)
 .
\end{equation}
These results indicate   the global $SU(3)$ {\it color symmetry preservation}, i.e., no specific direction in color space.  
This is expected, since the Yang-Mills theory should respect the global gauge symmetry, i.e., color symmetry, even after imposing the Landau gauge.

\begin{figure}[t]
\begin{center}
\includegraphics[height=5cm,width=8.0cm]{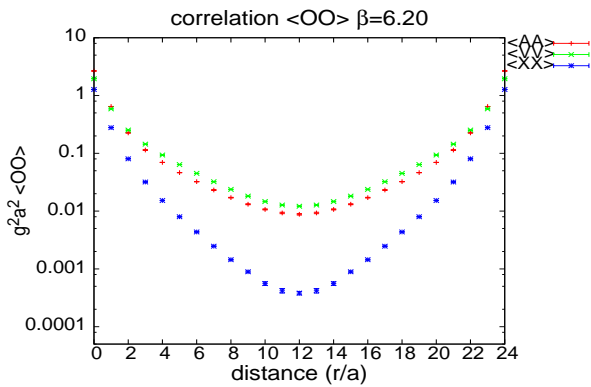}
\end{center}
\caption{Field correlators as functions of the distance $r:=|x|$ (from above to below)
$\langle  \mathscr{V}_\mu^A(x) \mathscr{V}_\mu^A(0) \rangle$, 
$\langle  \mathscr{A}_\mu^A(x) \mathscr{A}_\mu^A(0) \rangle$, 
and
$\langle  \mathscr{X}_\mu^A(x) \mathscr{X}_\mu^A(0) \rangle$.
}
\label{fig:decomp-field-corr}
\end{figure}

Fig. \ref{fig:decomp-field-corr} shows correlators of new fields $\mathscr{V}$, $\mathscr{X}$, and original fields $\mathscr{A}$.
This result indicates the {\it infrared  dominance} of restricted correlation functions $\langle  \mathscr{V}_\mu^A(x) \mathscr{V}_\mu^A(0) \rangle$ in the sense that the  correlator of the variable $\mathscr{V}$ behaves just like the correlator  $\langle  \mathscr{A}_\mu^A(x) \mathscr{A}_\mu^A(0) \rangle$ of the original variable $\mathscr{A}$ and dominates  in the long distance, while the correlator $\langle  \mathscr{X}_\mu^A(x) \mathscr{X}_\mu^A(0) \rangle$ of $SU(3)/U(2)$ variable $\mathscr{X}$  decreases quickly in the distance $r$.  
For  $\mathscr{X}$,  at least, we can introduce a gauge-invariant mass term \begin{equation}
\frac12 M_X^2 \mathscr{X}_\mu^A \mathscr{X}_\mu^A 
 ,
\end{equation} 
since $\mathscr{X}$ transforms like an adjoint matter field under the gauge transformation. 
In view of this fact, we fit the data of the contracted correlator $\langle  \mathscr{X}_\mu^A(x) \mathscr{X}_\mu^A(0) \rangle$ using the ``massive" propagator for large $r:=|x|$:
\begin{align}
 \langle  \mathscr{X}_\mu^A(x) \mathscr{X}_\mu^A(0) \rangle 
&=   \int \frac{d^4k}{(2\pi)^4} e^{ikx} \frac{3}{k^2+M_X^2} 
\nonumber\\
&\simeq  {\rm const.} \frac{e^{-M_X r}}{r^{3/2}} 
  .
\end{align}
Then the naively estimated ``mass" $M_X$ of $\mathscr{X}$ is 
\begin{equation}
 M_X = 2.409 \sqrt{\sigma_{\rm phys}} = 1.1 {\rm GeV} 
  .
\end{equation} 
 This value should be compared with the result in MA gauge \cite{SAIIMT02}. 
For more preliminary results of numerical simulations, see \cite{Shibata-lattice2009} for various properties of magnetic monopoles of SU(2),  \cite{Shibata-lattice2007} for the maximal option of SU(3) and \cite{Shibata-lattice2010,Shibata-lattice2008} for the minimal one of SU(3).

\section{Conclusion and discussion}

In this paper, we have given a new reformulation called the minimal option of the $SU(N)$ Yang-Mills theory in the continuum and on the lattice by elaborating previous works to provide one with an efficient framework to study quark confinement originating from magnetic monopoles defined  in the gauge-independent manner.

In fact, we have given first numerical evidences that 
 non-Abelian magnetic monopoles defined in a gauge-invariant way in this paper are dominant for confinement of fundamental quarks in SU(3) Yang-Mills theory.
By using the gauge invariant magnetic current $k$, we have extracted just the $U(1)$ part of  the stability group $U(N-1) \simeq SU(N-1) \times U(1)$ for the non-Abelian magnetic monopole associated with quarks in the fundamental representation, which is consistent with the consideration of the Homotopy group (\ref{SU3-mm}). 
This $U(1)$ part is enough to extract the dominant part of the Wilson loop average. 

This fact suggests that the non-Abelian  dual superconductivity caused by condensation of non-Abelian magnetic monopoles could be a mechanism for quark confinement in SU(3) Yang-Mills theory.
However, in order to establish the non-Abelian dual superconductivity in SU(3) Yang-Mills theory in this sense,  we must confirm that the spontaneous breaking of the dual $U(1)$ symmetry is associated with condensation of the non-Abelian magnetic monopoles obtained in this paper.  
For this purpose, we need to specify how the relevant low-energy effective theory of the $SU(3)$ Yang-Mills theory looks like, although it is believed to be a dual Ginzburg-Landau model based on the dual superconductor picture for quark confinement.

In order to see directly the non-Abelian nature of magnetic monopoles  defined in this paper, it will be necessary to study how the interaction among the non-Abelian magnetic monopoles is described in the short distance where the internal non-Abelian $SU(2)$ degrees of freedom in $U(2) \simeq SU(2) \times U(1)$ other than $U(1)$ are expected to become relevant.  
Moreover, it will be interesting to study the contribution of non-Abelian magnetic monopoles to gluon confinement. 
In the framework of the reformulation using new variables, a first step in this direction was taken recently in \cite{Kondo11} for $SU(2)$ gauge group. 
Extending this work to the $SU(3)$ gauge group will be a main subject of subsequent works.

{\it Acknowledgements}\ ---
This work is  supported by Grant-in-Aid for Scientific Research (C) 21540256 from Japan Society for the Promotion of Science (JSPS) 
and also by the JSPS Grant-in-Aid for Scientific Research (S) \#22224003. 
The numerical calculations are supported by the Large Scale Simulation Program No.09-15 (FY2009) and No.09/10-19 (FY2009-2010) of High Energy Accelerator Research Organization (KEK).


\end{document}